\documentclass{emulateapj}

\begin{document}
\title{Bimodal Infrared Colors of the M87 Globular Cluster System: Peaks in the Metallicity Distribution\altaffilmark{1}}

\author{Arunav Kundu\altaffilmark{2}, \& Stephen E. Zepf\altaffilmark{2}}

\altaffiltext{1} {Based on observations made with the NASA/ESA Hubble Space Telescope which is operated by
    the Association of Universities for Research in Astronomy, Inc., under NASA contract
    NAS 5-26555.}

\altaffiltext{2}{ Physics \& Astronomy Department, Michigan State University, East Lansing, MI 48824; e-mail: akundu, zepf @pa.msu.edu}

\begin{abstract}

The globular cluster (GC) systems of many galaxies reveal bimodal optical color
 distributions. Based on stellar evolutionary models and the bimodal colors and
 metallicities of Galactic GCs this is thought to reflect an underlying bimodal
 metallicity distribution. However, stars at many different phases of stellar
 evolution contribute to optical light. The I-H color is a much cleaner tracer
 of metallicity because it primarily samples the metallicity sensitive giant
 branch. Therefore, we use deep HST-NICMOS H, and WFPC2 optical observations, of
 M87 GCs to study their metallicity distribution. The M87 clusters are bimodal
 in I-H, for which there is no known physical explanation other than a bimodal
 metallicity distribution. Moreover, the two modes defined by the B-I and I-H
 colors are comprised of roughly the same two sets of objects, confirming that
 optical colors also primarily trace the metallicity. This is inconsistent with
 a recent suggestion based on one  model of metallicity effects on the
 horizontal branch that bimodality arises from an underlying unimodal
 metallicity distribution due to a specific color-metallicity relation.  We also
 find no discernable variation in the peak colors of the M87 GCs out to $\sim$75
 kpc due to the declining ratio of red-to-blue GCs, as implied by this model.
 Similarly, there is no  evidence that the bimodal peaks are bluer for systems
 with large blue-to-red GC ratio.  Our observations confirm that the primary
 cause of bimodality in cluster systems is an underlying bimodal metallicity
 distribution, and not the specific color-metallicity relationship defined by
 this horizontal branch model.

\end{abstract}


\section{Introduction}
 It has long been recognized that globular clusters (GC) provide unique insights into the formation histories of their host galaxies. The specific age and metallicity associated with each of these bright, simple stellar systems makes it much easier to interpret their color and spectrum than the typically complex, luminosity weighted emission from multiple generations of stars that constitute the underlying galaxy. The ensemble of GCs around a galaxy thus provides a unique set of fingerprints that can help decipher the formation history of its host galaxy.

Largely due to advances in instrumentation there has been an explosion of GC studies in recent times. Among the most striking discoveries is that a majority of globular cluster systems are bimodal in optical colors (e.g. Gebhardt \& Kissler-Patig 1999; Kundu \& Whitmore 2001 [KW01]). While most studies have concentrated on GC rich elliptical galaxies, the Milky Way and M31 reveal similar color bimodality (e.g. Barmby et al. 2000), albeit with a different ratio of blue to red GCs. 

For relatively old ages (more than a few Gyrs) stellar evolution models predict that broad-band optical colors are primarily sensitive to the metallicity of a stellar system (e.g. Bruzual \& Charlot 2003; Maraston 2005). The bimodal color distributions of GCs in the Milky Way and M31 are also known to correspond to two distinct metallicity peaks (Ashman \& Zepf 1998; Barmby et al. 2000; Perrett et al. 2002). Thus the bimodality in GC colors has typically been interpreted as a bimodality in the metallicities of the clusters, with corresponding implications on various galaxy formation scenarios (e.g. Ashman \& Zepf 1992; Cote et al. 1998; Beasley et al. 2002). Follow-up studies have shown that the blue and red GCs in ellipticals typically have different spatial (e.g. Geisler, Lee,\& Kim 1996) and kinematic (e.g. Zepf et al. 2000; Cote et al. 2003) properties, indicating (at least) two distinct populations.  However, in a recent paper Yoon, Yi, \& Lee (2006; hereafter YYL06) claim that while the Milky Way has a more ``complex" origin than ellipticals, and hence a bimodal metallicity distribution, the bimodal color peaks in elliptical galaxies  represent a unimodal metallicity distribution. YYL06 suggest that a metallicity dependent change in the horizontal branch structure of GCs results in an inflection point in the optical color vs. [Fe/H] relation, causing an artificial dip in the broad band optical colors.

\begin{figure}
\includegraphics[width=8cm]{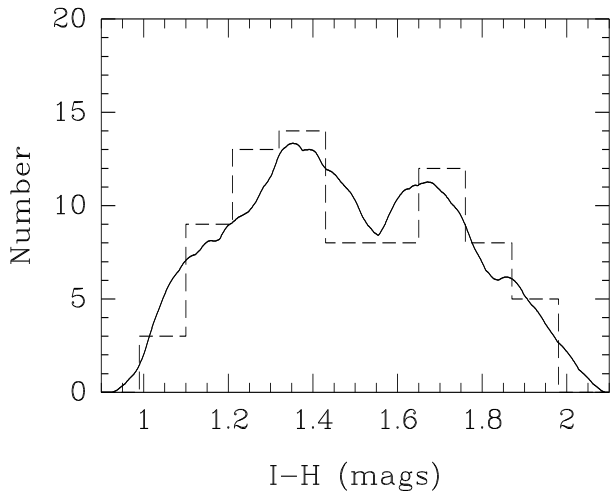}
\caption{  The dashed line plots a histogram of I-H colors of GCs in M87 while the smooth line traces the fixed width Epanechnikov kernel smoothed density distribution.  KMM mixture modelling analysis on the unbinned data gives a greater than 98\% probability that the distribution is bimodal rather than unimodal.  }
\end{figure}

Near-infrared observations provide a means to directly measure the metallicity of GCs, in particular the red GCs that are preferentially located in the high surface brightness inner regions of ellipticals. While optical colors are sensitive to the main sequence, as well as the horizontal and giant branches, near-IR colors are almost exclusively determined by the red giant branch and are hence sensitive estimators of the metallicity of a stellar population. Advances in IR imaging now allow such direct measurements of GCs in the inner regions of galaxies (e.g. Puzia et al. 2002; Hempel et al. 2007). In this Letter we analyze the deepest near-IR images of a globular cluster system to date. Using the HST - Near Infrared Camera and Multi-Object Spectrograph (NICMOS) camera, we test whether the metallicity distribution in M87, the giant elliptical in the center of the Virgo Cluster, follows the bimodal optical color distribution, or is inherently unimodal. We also test the YYL06 suggestion that their model can explain the radial distribution of GC colors. In principle optical spectroscopy can also probe the metallicity distribution of GCs. However, the small sample sizes of such studies
 (e.g. Puzia et al. 2005; Strader et al. 2005) and difficulties in observing metal-rich red GCs that are generally concentrated in the bright inner regions of the galaxies limits their usefulness. This is apparent in Fig 2 of YYL06 (panels I \& M) which show few GCs with reliable metallicities past the inflection point in the YYL06 model. Moreover, optical spectroscopy is also sensitive to horizontal branch structure (e.g. Cohen et al. 2003).

 \section{Near-IR Colors and the Metallicity Distribution of M87 Globular Clusters }

The inner region of M87 was observed for 11,007.5 secs in the H-band (F160W) with the NICMOS-NIC3 imager on 12$^{th}$ and 13$^{th}$ Jan 2006 (GO-10529 PI - Kundu). These observations were sub-pixel dithered at four positions. This places the center of each GC  at different locations with respect to the center of a pixel, thus reducing the effects of intrapixel sensitivity variations in the NIC3 array (Xu \& Mobasher 2003).  Using the drizzle algorithm (Fruchter \& Hook 2002) we reconstructed high resolution images, alleviating the effect of the undersampled 0.2'' NIC3 pixels. Since the count dependent non-linearity  is only 0.02 mag per dex in this configuration (de Jong 2006) and the GCs are superposed on the bright galaxy background (reducing the contrast), we did not correct for this. 

The GCs in this region of M87 have also been studied in the V (F606W) and I (F814W) bands by Waters et al. (2006) using deep HST-WFPC2 observations. Using the GC lists kindly provided to us by Chris Waters we measured the H-band magnitudes of the 191 GC candidates within our field in a manner similar to Kundu et al. (2005). We note that {\it all} photometry in this paper is reported in the Vegamag system and corrected  for foreground reddening using Schlegel, Finkbeiner, \& Davis (1998).  In Figure 1 we plot the I-H distribution of the 80 GCs with uncertainties less than 0.1 mag.  It is particularly important to select only the high signal-to-noise, and hence bright, sources because the faint, and therefore low mass GCs, are likely to have only a handful of red giant branch stars. Stochastic variations in the locations of red giants in the sparse red giant branch of a low mass, and especially a metal-rich, GC can introduce additional scatter in near-IR colors (Bruzual \& Charlot 2003; Cohen et al. 2007). 

Both the histogram of the I-H color distribution and the fixed width Epanechnikov kernel smoothed density distribution reveal that the color distribution is bimodal, and hence inconsistent with the unimodal metallicity (and therefore I-H color) distribution suggested by YYL06. KMM mixture modelling tests (Ashman, Bird \& Zepf 1994) confirm that the distribution is bimodal at greater than 98\% confidence, with peaks at I-H of 1.30 and 1.70 and a dividing color of I-H=1.51.

\begin{figure}
\includegraphics[width=8cm]{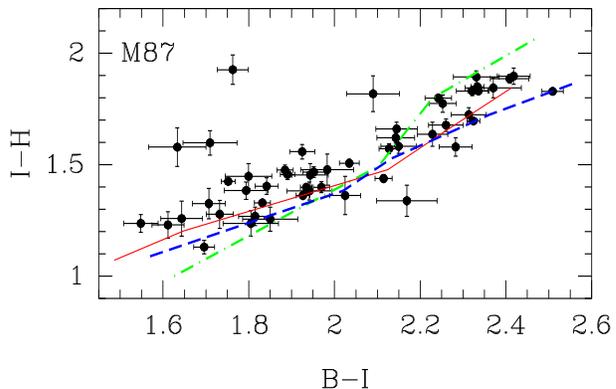}
\caption{ The B-I vs. I-H distribution of globular clusters in the central region of M87. The distribution is consistent with the predictions of the stellar evolution models of Bruzual \& Charlot (2003) [dot-dash line], Maraston (2005) [solid line], and Anders \& Fritze-v. Alvensleben (2003) [dashed line] and does not show evidence of the inflection point due to horizontal branch structure suggested by Yoon et al. (2006). The gap in the B-I vs I-H distribution near B-I 2.05 and I-H 1.5 reveals that the bimodality in the optical and near-IR colors arises from a lack of GCs with intermediate metallicities and colors. The apparently large blue to red GC ratio is a selection effect due to the shallower optical B data.  }
\end{figure}

This field of view has also been imaged with the HST-WFPC2 camera in the B (F450W) and I (F814W) bands on 11$^{th}$ May 1999, 15$^{th}$ April 2001, 18$^{th}$ May 2001, and 13$^{th}$ July 2002 in a program to monitor the M87 jet. These observations totalling 1280s in F450W  and 1600s in F814W were reduced and analyzed in the manner described in KW01. Fifty seven GCs were detected within the NIC3 field of view. The F450W and F814W  observations were converted to B and I using color transformations provided to us by Jon Holtzman (private communication).

\begin{figure*}[!ht]
\includegraphics[angle=-90, width=14cm]{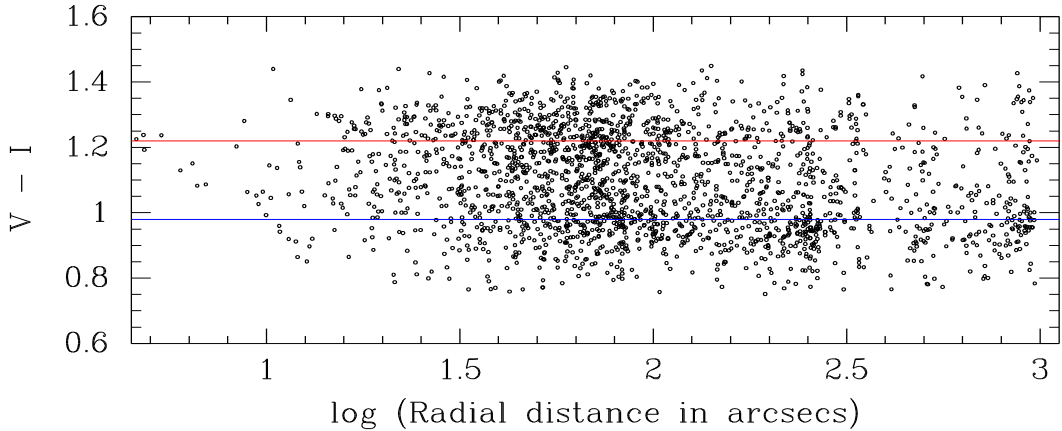}
\caption{ The radial distribution of V-I colors of GCs in M87. Note that the spatial coverage is not uniform at all radii. The figure shows that the red GCs are more spatially concentrated than the blue ones. There is no evidence that the peaks of the red and blue GCs change with distance (solid lines). This is at odds with the Yoon et al. (2006) scenario which suggests that both the red and blue peaks should drift to bluer colors at larger radii because of the decrease in the mean metallicity.  }
\end{figure*}

In Fig 2 we plot the I-H vs B-I colors of the of the M87 GCs with an uncertainty less than 0.1 in each color.  It is clear that there are two distinct populations with a gap at B-I$\approx$2.05 and I-H$\approx$1.5. The fact that roughly the same set of objects define the two modes in the  I-H and the B-I color distributions  shows that the reason for the bimodality in the optical colors such as B-I is primarily the metallicity, and not some other effect such as variations in the horizontal branch. We overlay the isochrones for old (13-15 Gyrs) globular clusters from the stellar population models of Bruzual \& Charlot (2003), Maraston (2005), and Anders \& Fritze-v. Alvensleben (2003). The color metallicity curves derived by YYL06 are not in the public domain so we could not overlay them on Fig 2. There are differences in the detail between the various models due to the different stellar evolution tracks used and the treatment of late stages of stellar evolution (Maraston 2005 and references within). If instead of the blue horizontal branch Maraston (2005) model plotted in Fig 2 the red horizontal branch model is plotted the curves lie almost identically on the Anders tracks suggesting that the uncertainties in model calibrations are comparable to any effect due to horizontal branch changes. Deep near-IR datasets such as this one will be an effective way to test and refine these models. However, Fig 2 shows that there is no need to invoke an inflection point in the color metallicity (or I-H) curve, as suggested by YYL06. Hence, the bimodality in the optical colors of ellipticals is primarily indicative of the bimodal metallicity of the underlying globular clusters.

 \section{The Spatial Distribution of Globular Clusters and Predictions from the Horizontal Branch Models }

The peak color of the red globular cluster system of various galaxies is known to be correlated with the mass of the host galaxy (e.g. KW01; Larsen et al. 2001; Peng et al. 2006) such that more massive galaxies have redder peaks. Some studies also suggest a similar, albeit weaker, trend for the peaks of the metal-poor GCs (however see Kundu \& Zepf 2007, in preparation). YYL06 explain this by the location of the peak of their unimodal metallicity distribution. According to this explanation as the mean metallicity of the GCs increases with galaxy mass there is always a lack of GCs in the fixed color range near the inflection point, and hence the peak colors of both the red and blue GCs become redder. A direct consequence of this theory is that the ratio of red to blue GCs governs the location of the red and blue peaks, i.e. as the fraction of red GCs in a system increases the peaks of the red and blue GC subpopulations should drift redder. YYL06 also suggest that the apparent difference in the spatial distributions of the red and blue GCs in a particular galaxy is due to the decreasing mean metallicity of the GCs with galactocentric distance. Thus their theory implies that the peaks of both subsystems should trend bluer at larger galactocentric distances, as the fraction of blue GCs increases.

We have previously analyzed WFPC2 V and I images of M87 roughly covering the entire radial range from the center of M87 to $\approx$15' (Vesperini et al. 2003). Fig 3 plots the V-I colors of the GCs as a function of galactocentric distance. The bimodal nature of the GC colors is clearly visible. It is also clear from the figure that the red GCs are more centrally concentrated than the blue ones. Hence, the blue to red fraction increases, with increasing galactocentric distance. However, the peaks of the red and blue subsystems do not appear to shift consistently to bluer colors at larger distances, as implied by YYL06.

In order to quantify the shift in the peak colors we performed KMM tests on the color distribution for each pointing. In Fig 4 we plot the locations of the red and blue peaks as a function of the ratio of blue to red clusters. Each set of triangles at a particular N$_{blue}$/N$_{red}$ ratio marks the peak colors of GCs in a particular field of view, and hence represents a certain  galactocentric distance (see Fig 3). The peaks appear to be roughly constant at all points, and hence all galactocentric radii, irrespective of the ratio of red to blue GCs at that location. The redder set of peak colors at N$_{blue}$/N$_{red}$ about 1.5 corresponds to a field that is $\approx$4' from the center, while the bluer set corresponds to another set of observations at $\approx$8'. Interestingly the field at $\approx$8' is centered on NGC 4486A, a low mass companion of M87. Clearly the unexpectedly low ratio of blue to red GCs at this large galactocentric distance from M87 is due to the contribution of NGC 4486A GCs. This suggests that some other process associated with the luminosity of the host galaxy is the cause of this difference. Thus horizontal branch variations cannot explain the (lack of) radial variations in the peak colors of the M87 GC system, whereas it is easily explained by the different spatial distribution of two chemically (and dynamically) distinct populations.

\begin{figure}
\includegraphics[angle=0, width=8cm]{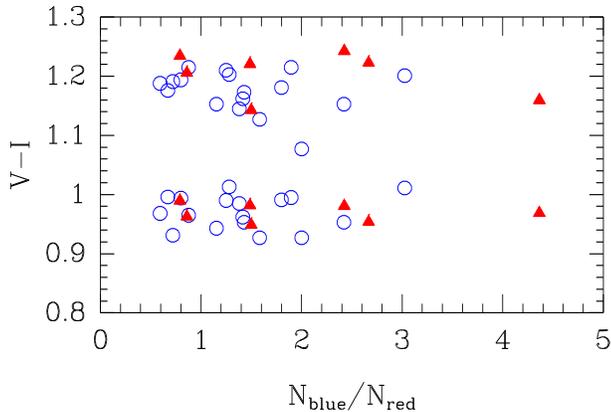}
\caption{The circles plot the locations of the red and blue peaks of the bimodal GC systems identified in Kundu \& Whitmore (2001) vs. the ratio of blue to red GCs. There is no evidence that peaks of the distributions become bluer as the fraction of blue GCs increases as expected by the Yoon et al. (2006) scenario. The triangles mark the location of the peaks of the red and blue GCs at various radii in M87 (see Fig 2), and shows no evidence of the Yoon et al. (2006) correlation. See text for details.  }
\end{figure}

For comparison we also plot the peaks of the red and blue GCs for the galaxies with confirmed, or likely, bimodality in KW01 in Fig 4. Note that we have corrected the KW01 data with the updated zeropoints from the HST data handbook and Schlegel et al. (1998) reddenings. Fig 4 shows that the peaks of the red and blue GCs in various galaxies do not drift to bluer colors as the fraction of blue GCs increases. 
We have found that other published bimodality data (e.g. Peng et al. 2006) show a similar lack of correlation between the blue-to-red ratio and the peak colors.
 The lack of a clear trend between the N$_{blue}$/N$_{red}$ and the peak colors,  despite the correlation of the red peaks, and the much weaker correlation of the blue peaks  with galaxy mass (KW01; Larsen et al. 2001; Peng et al. 2006) suggests that other factors contribute to a larger scatter in the galaxy to galaxy variation. Outside the constraints of the YYL06 model N$_{blue}$/N$_{red}$ can change due to a variety of factors such as different spatial distributions of the two subsystems, different dynamical destruction effects, different enrichment histories of the galaxies, etc. We note for example that small HST field of view studies such as Peng et al. (2006) only sample a fraction of the GC systems of large galaxies such as M87, and preferentially the red ones in the inner regions, but likely the entire GC system of the faintest galaxies in their sample. Thus the apparent N$_{blue}$/N$_{red}$ fraction has galaxy mass bias.  Regardless of the choice of galaxies, and cluster subsamples, as long as the N$_{blue}$/N$_{red}$ ratio spans a range of values (as the data in Fig 4 does) YYL06 implies that the peak must change in concert.
Fig 4 is inconsistent with such a prediction.
 The galaxy with the largest ratio of blue to red GCs in Fig 4 is NGC 4406, a bright elliptical in Virgo. The location of the peaks in this galaxy is not changed even though the high blue to red ratio has been confirmed by many studies (e.g. KW01; Peng et al. 2006). This is further evidence that the location of the peak colors of GCs in early type galaxies is not primarily governed by the superposition of horizontal branch properties on a unimodal metallicity distribution but by an inherently bimodal metallicity distribution.

 \section{Discussion and Conclusions}

Using deep HST-NICMOS near-IR, and WFPC2 optical observations of GCs in M87 we show that a bimodal distribution of I-H colors is strongly preferred over a unimodal one. Since these bands are not sensitive to the horizontal branch structure of GCs, our observations reveal that the metallicity distribution of the GCs in this intensely studied elliptical galaxy is bimodal, and not unimodal as suggested in the YYL06 scenario. Roughly the same set of GCs fall into the red and blue subsystems, irrespective of the whether the two populations are separated by the IR or optical colors, confirming that the two optical subsystems are distinguished by their metallicity and not their horizontal branch structure. The YYL06 model also suggests that peak colors of the red and blue GCs should drift towards bluer colors at larger galactocentric radii as red clusters become scarcer. However we see no evidence of a galactocentric distance dependent change in the peak colors of M87 GCs in Figs 3 and 4. This further suggests that the primary cause of the bimodality in the optical colors of GCs in ellipticals is not an inflection point in the color metallicity relation of a unimodal GC metallicity distribution due to changes in horizontal branch structure but two underlying metallicity populations. In a recent spectroscopic analysis of NGC 4472 GCs, Strader, Beasley, \& Brodie (2007) reach a similar conclusion. The apparent inflection in the color metallicity distribution in Fig 1. of YYL06 may arise from uncertainties in the relative calibrations of the heterogeneous data sets combined for that plot.

YYL06 argue that the clear bimodal metallicity in the Milky Way is due to the ``complicated" history of our spiral galaxy. However, complicated (major) merger histories are found to produce elliptical galaxies both in simulations (e.g. Barnes \& Hernquist 1992), as well as observations (Schweizer \& Seitzer 1992). On the other hand
the disks of spiral galaxies are readily destroyed by major mergers (e.g. Hernquist 1992) and thus spiral galaxies would appear to be those that have a more sedate formation history.

While it is clear that horizontal branch stars contribute some fraction of the optical light, especially in the bluest filters, it is not at all clear that the horizontal branch structure faithfully follows the mean metallicity
of the GCs as assumed by YYL06. Based on the evidence at hand it is clear that there are GCs with at least two different metallicities in M87.  Since the B-I colors trace these same populations the optical colors primarily trace an underlying bimodal metallicity distribution. By extension the bimodal optical color distributions of GCs in ellipticals trace an underlying bimodal metallicity distribution.

We gratefully acknowledge support from NASA-LTSA grant and NAG5-12975,  and HST grants AR-09208 and GO-10529. We thank Katherine Rhode for useful discussions and Chris Waters for providing optical source lists.


\begin{thebibliography}{}
\bibitem[Anders \& Fritze-v.~Alvensleben(2003)]{2003A&A...401.1063A} 
Anders, P., \& Fritze-v.~Alvensleben, U.\ 2003, \aap, 401, 1063 

\bibitem[A94]{A94} Ashman, K. M., Bird, C. M., \& Zepf, S. E. 1994, \aj, 108, 2348

\bibitem[Ashman \& Zepf(1992)]{1992ApJ...384...50A} Ashman, K.~M., \& Zepf, 
S.~E.\ 1992, \apj, 384, 50 

\bibitem[AZ98]{AZ98} Ashman, K. M., \& Zepf, S. E. 1998, "Globular Cluster Systems", (Cambridge: Cambridge Univ.\ Press)
\bibitem[Barmby et al.(2000)]{2000AJ....119..727B} Barmby, P., Huchra, 
J.~P., Brodie, J.~P., Forbes, D.~A., Schroder, L.~L., \& Grillmair, C.~J.\ 
2000, \aj, 119, 727 
\bibitem[Barnes \& Hernquist(1992)]{1992ARA&A..30..705B} Barnes, J.~E., \& 
Hernquist, L.\ 1992, \araa, 30, 705 
\bibitem[Beasley et al.(2002)]{2002MNRAS.333..383B} Beasley, M.~A., Baugh, 
C.~M., Forbes, D.~A., Sharples, R.~M., \& Frenk, C.~S.\ 2002, \mnras, 333, 
383 
\bibitem[BC00]{BC00} Bruzual, A. G., \& Charlot, S. 2003, MNRAS, 344, 1000
\bibitem[Cohen et al.(2003)]{2003ApJ...592..866C} Cohen, J.~G., Blakeslee, 
J.~P., \& C{\^o}t{\'e}, P.\ 2003, \apj, 592, 866
\bibitem[Cohen et al.(2007)]{2007AJ....133...99C} Cohen, J.~G., Hsieh, S., 
Metchev, S., Djorgovski, S.~G., \& Malkan, M.\ 2007, \aj, 133, 99 
\bibitem[Cote et al.(1998)]{1998ApJ...501..554C} Cote, P., Marzke, R.~O., 
\& West, M.~J.\ 1998, \apj, 501, 554
 \bibitem[C{\^o}t{\'e} et al.(2003)]{2003ApJ...591..850C} C{\^o}t{\'e}, P., 
McLaughlin, D.~E., Cohen, J.~G., \& Blakeslee, J.~P.\ 2003, \apj, 591, 850 
\bibitem[X03]{X03} de Jong 2006, NICMOS Instrument Science Report 2006-003 
\bibitem[F02]{F02} Fruchter, A. S., \& Hook, R. N. 2002, PASP, 114, 144


\bibitem[Geisler et al.(1996)]{1996AJ....111.1529G} Geisler, D., Lee, 
M.~G., \& Kim, E.\ 1996, \aj, 111, 1529 
\bibitem[Gebhardt \& Kissler-Patig(1999)]{1999AJ....118.1526G} Gebhardt, 
K., \& Kissler-Patig, M.\ 1999, \aj, 118, 1526
\bibitem[H07]{H07} Hempel, M., Zepf, S. E., Kundu, A., Geisler, D., \& Maccarone, T. J. 2006, \apj, in press 
\bibitem[Hernquist(1992)]{1992ApJ...400..460H} Hernquist, L.\ 1992, \apj, 
400, 460 
\bibitem[Kundu et al.(2005)]{2005ApJ...634L..41K} Kundu, A., et al.\ 2005, 
\apjl, 634, L41 
\bibitem[KW01]{KW01} Kundu, A., \& Whitmore, B. C. 2001, \aj, 121, 2950

\bibitem[Maraston(2005)]{2005MNRAS.362..799M} Maraston, C.\ 2005, \mnras, 
362, 799 
\bibitem[Peng et al.(2006)]{2006ApJ...639...95P} Peng, E.~W., et al.\ 2006, 
\apj, 639, 95 
\bibitem[Perrett et al.(2002)]{2002AJ....123.2490P} Perrett, K.~M., 
Bridges, T.~J., Hanes, D.~A., Irwin, M.~J., Brodie, J.~P., Carter, D., 
Huchra, J.~P., \& Watson, F.~G.\ 2002, \aj, 123, 2490 
\bibitem[Puzia et al.(2005)]{2005A&A...439..997P} Puzia, T.~H., 
Kissler-Patig, M., Thomas, D., Maraston, C., Saglia, R.~P., Bender, R., 
Goudfrooij, P., \& Hempel, M.\ 2005, \aap, 439, 997 

\bibitem[P02]{P02} Puzia, T., Zepf, S. E.,  Kissler-Patig, M., Hilker, M., Minniti, D., Goudfrooij, P. 2002, A\&A, 391, 453

\bibitem[S98]{S98} Schlegel, D. J., Finkbeiner, D. P., \& Davis, M. 1998, ApJ, 500, 525

\bibitem[Schweizer \& Seitzer(1992)]{1992AJ....104.1039S} Schweizer, F., \& 
Seitzer, P.\ 1992, \aj, 104, 1039 
\bibitem[Strader et al.(2007)]{Strader et al.} Strader, J., Beasley, M.~A.\, \& Brodie, J.~P.  2007, \aj, in press

\bibitem[Strader et al.(2005)]{2005AJ....130.1315S} Strader, J., Brodie, 
J.~P., Cenarro, A.~J., Beasley, M.~A., \& Forbes, D.~A.\ 2005, \aj, 130, 
1315
\bibitem[Vesperini et al.(2003)]{2003ApJ...593..760V} Vesperini, E., Zepf, 
S.~E., Kundu, A., \& Ashman, K.~M.\ 2003, \apj, 593, 760 
\bibitem[Waters et al.(2006)]{2006ApJ...650..885W} Waters, C.~Z., Zepf, 
S.~E., Lauer, T.~R., Baltz, E.~A., \& Silk, J.\ 2006, \apj, 650, 885 
 \bibitem[X03]{X03} Xu, C., \& Mobasher, B. 2003, NICMOS Instrument Science Report 2003-009 

\bibitem[Yoon et al.(2006)]{2006Sci...311.1129Y} Yoon, S.-J., Yi, S.~K., \& 
Lee, Y.-W.\ 2006, Science, 311, 1129 

\bibitem[Zepf et al.(2000)]{2000AJ....120.2928Z} Zepf, S.~E., Beasley, 
M.~A., Bridges, T.~J., Hanes, D.~A., Sharples, R.~M., Ashman, K.~M., \& 
Geisler, D.\ 2000, \aj, 120, 2928 

\end{thebibliography}
\end{document}